# The Effects of Powertrain Mechanical Response on the Dynamics and String Stability of a Platoon of Adaptive Cruise Control Vehicles

L. C. Davis
10244 Normandy Dr.
Plymouth, MI 48170

ABSTRACT

The dynamics of a platoon of adaptive cruise control vehicles is analyzed for a general mechanical response of the vehicle's powertrain. Effects of acceleration-feedback control that were not previously studied are found. For small acceleration-feedback gain, which produces marginally string-stable behavior, the reduction of a disturbance (with increasing car number $n$) is found to be faster than for the maximum allowable gain. The asymptotic magnitude of a disturbance is shown to fall off as $erf\left(\frac{ct.}{\sqrt{n}}\right)$ when $n \to \infty$. For gain approaching the lower limit of stability, oscillations in acceleration associated with a secondary maximum in the transfer function (as a function of frequency) can occur. A frequency-dependent gain that reduces the secondary maximum, but does not affect the transfer function near zero frequency, is proposed. Performance is thereby improved by elimination of the undesirable oscillations while the rapid disturbance reduction is retained.

**Keywords:** adaptive cruise control, string stability, mechanical response, platoon dynamics

1. Introduction

The dynamics, in particular the string stability, of a platoon of adaptive cruise control (ACC) vehicles has been studied by numerous authors (see Sheikholeslam and Desoer, 1990; Ioannou and Chien, 1993; Swaroop et al., 1994; Swaroop and Hedrick, 1996; Liang and Peng, 1999; Liang and Peng, 2000; Treiber and Helbing, 2001; Li and Shrivastava, 2002; Bareket et al., 2003; Yi and Horowitz, 2006). Interest in ACC vehicles is strong because they provide higher flow rates and have a lower probability to form jams than conventional vehicles driven by people. A major factor in the improved performance is the elimination of human reaction time. Likewise, mixing ACC vehicles in with manually driven vehicles has been shown to improve stability and increase traffic flow (VanderWerf et al., 2002; Bose and Ioannou, 2003a; Bose and Ioannou; 2003b; Davis, 2004; Ioannou and Stefanovic, 2005; Kerner, B. S., 2005; Jiang and Wu, 2006; Davis, 2007; Jiang et al., 2007; Kesting et al., 2008; Yuan et al., 2009).

In recent years the effects of a time delay and a first-order time constant (associated with the vehicle's mechanical response) on the stability of ACC platoons have been discussed (Zhou and Peng, 2005; Orosz et al., 2010; Orosz et al., 2011; Davis, 2012). At low speeds the mechanical response can be slow enough to be a factor in stability. At higher speeds, the faster response is not so important unless small headways are desired, in which case the same considerations apply as at low speed. The purpose of the present work is to examine the effects of mechanical response more generally and in more detail than what has been reported previously.

In Sec. 2, a formalism is presented for calculating the dynamics of a platoon of ACC vehicles with a general response function that is not limited to a description based on an explicit delay and a first-order time constant. Sec. 3 contains simulations for a realistic response function modeled after the torque response to a step-function change in throttle angle (Stefanopoulou and Kolmanovsky, 1999). The rate at which a disturbance (such as that caused by a brief acceleration of the leading vehicle) decreases with increasing car number (from front to rear) is determined in Sec. 4. A frequency-dependent gain for acceleration-feedback control is proposed and evaluated in Sec. 5. Conclusions are drawn in Sec. 6.

2. General response

The control algorithm for the desired acceleration of each vehicle of a platoon is taken to be (Zhou and Peng, 2005)

$$a^d(t) = \frac{\alpha}{h}[x_{lead}(t) - x(t) - D] - \alpha v(t) + k[v_{lead}(t) - v(t)] - \xi a(t). \tag{1}$$

The subscript "lead" designates the preceding vehicle and the parameters in Eq. (1) are the sensitivity $\alpha$, the headway-time constant $h$, the coefficient of relative-velocity feedback $k$, and the acceleration-feedback gain $\xi$.

If the desired acceleration were a step function

$$a^d(t) = 0, t < 0 \tag{2a}$$

$$= 1, t \geq 0, \tag{2b}$$

the resulting acceleration would be $g(t)$, which has the properties

$$g(t) = 0, t \leq 0 \tag{3}$$

and

$$g(t) \to 1, t \to \infty. \tag{4}$$

For an arbitrary desired acceleration the acceleration is therefore

$$a(t) = \int_0^t \frac{dg(t')}{dt'} a^d(t - t') dt'. \tag{5}$$

To determine the transfer function $G(\omega)$ in Fourier space let

$$\Lambda(\omega) = 1 - i\omega \int_0^\infty e^{-i\omega t} [1 - g(t)] dt. \tag{6}$$

Then the control algorithm Eq. (1) and the definition Eq. (6) give

$$G(\omega) = \frac{\frac{\alpha}{h} + ik\omega}{\frac{\alpha}{h} - \omega^2 [\xi + \Lambda(\omega)^{-1}] + i\omega(\alpha + k)}. \tag{7}$$

### 3. Simulations

To perform simulations I introduce discrete-time control for convenience. Let the desired acceleration now be given by

$$a^{desired}(t) = a_{d\mu}, \ \mu T \leq t < (\mu + 1)T. \tag{8}$$

$T$ is the update time, typically 0.01 s and $a_{d\mu} = a^d(\mu T)$ as given by Eq. (1). [$\mu$ is an integer.]

The acceleration is then

$$a(t) = \sum_{m=0}^\mu a_{dm} [g(t - mT) - g(t - mT - T)], \ \mu T \leq t < (\mu + 1)T. \tag{9}$$

Suppose a sufficiently accurate description of the response function is given by

$$g(t) = g_\mu, \ \mu T \leq t < (\mu + 1)T, \ 0 < \mu \leq N \tag{10a}$$

$$g(t) = 1, \ t > NT, \tag{10b}$$

then for $t < \mu T$

$$\int_{\mu T}^{t} dt' a(t') = (t - \mu T) \sum_{k=0}^{N} \Gamma_k a_{d\mu-k}. \tag{11}$$

If $\mu < N$, the upper value of k in the sum in Eq. (11) is $\mu$. Here

$$\Gamma_1 = g_1, \tag{12a}$$

$$\Gamma_k = g_{k+1} - g_k, \ k < N \tag{12b}$$

$$\Gamma_N = 1 - g_N. \tag{12c}$$

The velocity $v(t)$ and the position $x(t)$ of any vehicle are given by

$$v(\mu T + T) = v(\mu T) + T \sum_{k=0}^{N} \Gamma_k a_{d\mu-k}, \tag{13a}$$

$$x(\mu T + T) = x(\mu T) + v(\mu T)T + \tfrac{1}{2}T^2 \sum_{k=0}^{N} \Gamma_k a_{d\mu-k}. \tag{13b}$$

To perform simulations for a realistic $g(t)$, rather than a simple model with a first-order time constant and an explicit delay, I use the measured engine torque response curve for a step-function change in throttle angle (Stefanopoulou and Kolmannovsky, 1999). See Fig. 1. The acceleration response of the vehicle's powertrain can be no faster than the torque response. In fact, it will undoubtedly be slower because of other delays. If these delays were known, they could be easily included in the present analysis. Because the $g(t)$ pertains to only 750 rpm, I do simulations for low speeds. For all results reported, the leading vehicle (in front of the platoon) is initially moving at $v_0(t) = 5$ m/s until $t = 5$ s; then it accelerates for $T_a = 5$ s at $a_0(t) = 1$ m/s² and for $t > 10$ s maintains a constant velocity of 10 m/s. The parameters of the control algorithm Eq. (1) are given in Table 1. The cars in the platoon following the leading vehicle are numbered $n = 1, 2 \ldots$ from front to rear. At $t = 0$ all vehicles are moving at 5 m/s and are spaced 10 m apart (center-to-center distance). The vehicles are identical and all have the response function of Fig. 1.

The absolute square of the transfer function given by Eqs. (6) and (7) for the model $g(t)$ of Fig. 1 is shown in Fig. 2 for two values of the acceleration-feedback gain, $\xi = 0.82$ and 0.9. To have string stability it is well-known that $|G(\omega)| < 1$ for all $\omega > 0$, which is satisfied for both gains. However, the secondary peak in $|G(\omega)|^2$ at $\omega \approx 2.6$ rad/s for $\xi = 0.82$ is only slightly below 1. As a consequence, oscillations in the acceleration of vehicles occur following the main peak associated with accelerating from 5 m/s to 10 m/s. This effect is demonstrated in Fig. 3 for several vehicles. For example, the 50[th] car in the platoon has nearly a minute of oscillations reaching an amplitude of about 0.1 m/s², which is undesirable for ride comfort. Increasing the gain to 0.9 significantly reduces the oscillations as shown in Fig. 4. By the 50[th] car, only a small undershoot in the acceleration is apparent.

The requirement for a constraint on $|G(\omega)|$ in addition to the string-stability constraint has been discussed in a separate publication (Davis, 2013). The oscillations in acceleration due to the secondary peak in $|G(\omega)|$ are unique to analyses that include the mechanical response of the vehicles. With instantaneous response, instability is only associated with $|G(\omega)|$ rising above 1 near $\omega = 0$. In such a scenario, the main peak would grow in magnitude with increasing car number $n$. [It is also possible that the secondary peak can be greater than 1 (but $|G(\omega)|< 1$ near $\omega = 0$) in which case the oscillations increase in amplitude with increasing car number $n$.]

4. **Peak acceleration**

To approximately calculate the decrease in peak acceleration with increasing car number, I give the following analysis. Let the leading vehicle acceleration be

$$a_0(t) = 1, \ 0 < t < T_a \tag{14a}$$

$$= 0, \ otherwise. \tag{14b}$$

The Fourier transform is

$$A_0(\omega) = \frac{2}{\omega} e^{-i\omega \frac{T_a}{2}} \sin\left(\omega \frac{T_a}{2}\right). \tag{15}$$

For vehicle $n$ the acceleration is determined by

$$A_n(\omega) = [G(\omega)]^n A_0(\omega). \tag{16}$$

When $\omega \to 0$ the transfer function can be written as

$$G(\omega) = \exp(-\lambda(\omega) + i\theta(\omega)), \tag{17a}$$

$$\lambda(\omega) = t_2^2 \omega^2 + \cdots, \tag{17b}$$

$$\theta(\omega) = -h\omega + \cdots, \tag{17c}$$

where

$$t_2 = \left\{\frac{h}{2\alpha}[h(\alpha + 2k) - 2(1 + \xi)]\right\}^{\frac{1}{2}}. \tag{18}$$

The gain $\xi$ should not be so large as to make the RHS of Eq. (18) negative, otherwise the platoon would not be string stable (Liang and Peng, 1999; Davis, 2012).

The inverse Fourier transform of Eq. (17) gives

$$a_n(t) = \int_{-\infty}^{\infty} \frac{d\omega}{2\pi} e^{i\omega t} \exp\left(-nt_2^2 \omega^2 - inh\omega - i\omega \frac{T_a}{2}\right) \frac{2\sin\left(\frac{\omega T_a}{2}\right)}{\omega}. \tag{19}$$

At $t = nh + \frac{T_a}{2}$, which corresponds to the peak (maximum) acceleration,

$$a_n^{peak} = \frac{2}{\pi}\int_0^\infty \frac{d\omega}{\omega}\exp(-nt_2^2\omega^2)\sin\left(\frac{\omega T_a}{2}\right) \tag{20a}$$

$$= erf\left(\frac{T_a}{4t_2\sqrt{n}}\right). \tag{20b}$$

In Eq. (20b), $erf$ denotes the error function. This result is compared in Fig. 5 to the peak acceleration calculated in simulations. The agreement is good. Note that Eq. (20b) does not depend on $g(t)$ and thus would also apply to analyses in which the mechanical response is instantaneous.

5. **Frequency-dependent feedback**

Eqs. (18) and (20b) imply that there is a maximum allowable gain $\xi$ and that smaller gains give lower peak acceleration. If a smoother ride due to smaller acceleration is desired, then the gain should be no larger than necessary to eliminate oscillations caused by the secondary peak in $|G(\omega)|$. In this section I propose a frequency-dependent feedback scheme to produce better performance, although it requires additional memory and computation.

Normally the gain $\xi$ is a constant, however, the feedback can be generalized to include an $\omega$ dependence: $\xi(\omega)$. This implies that the acceleration-feedback term is of the form $\int_0^t \tilde{\xi}(t')a(t-t')dt'$ where

$$\tilde{\xi}(t) = \frac{1}{\pi}\int_0^\infty \xi(\omega)\cos(\omega t)\,dt. \tag{21}$$

A useful function for $\xi(\omega)$ is

$$\xi(\omega) = \xi_0 + C\omega^2 e^{-\omega^2/\omega_p^2} \tag{22}$$

where $\xi_0$, $C$ and $\omega_p$ are constants. For stability ($|G(\omega)| \leq 1$) it is required that [See Eq. (18).]

$$\xi_0 < \frac{h}{2}(\alpha + 2k) - 1. \tag{23}$$

The maximum value of the second term of Eq. (22) occurs at $\omega_p$ and the magnitude is

$$\xi_p = C\omega_p^2 e^{-1}. \tag{24}$$

The transform in the time domain of the second term of Eq. (22) is

$$\tilde{\xi}_p(t) = C_1\left(1 - \frac{t^2}{2\omega_p^2}\right)\exp(-\frac{t^2}{4\omega_p^2}) \tag{25}$$

where

$$C_1 = \frac{C\omega_p}{4\sqrt{\pi}}. \tag{26}$$

An example of Eq. (25) is shown in Fig. 6 for parameters $\xi_p = 0.3$ and $\omega_p = 5$ rad/s. In Fig. 7, the corresponding frequency-dependent acceleration gain, $\xi(\omega)$, with $\xi_0 = 0.8, \xi_p = 0.3$ and $\omega_p = 5$ rad/s is shown. The associated absolute square of the transfer function $|G(\omega)|^2$ is also depicted.

In the modification of the control algorithm Eq. (1), the acceleration–feedback now contains an additional term $\int_0^t \tilde{\xi}_p(t')a(t-t')dt'$ and the desired acceleration becomes

$$a^d(t) = \frac{\alpha}{h}[x_{lead}(t) - x(t) - D] - \alpha v(t) + k[v_{lead}(t) - v(t)] - \xi_0 a(t) - \int_0^t \tilde{\xi}_p(t')a(t-t')dt'. \tag{27}$$

For comparison to the results with the original algorithm, I show in Fig. 8 the absolute square of the transfer function, $|G(\omega)|^2$ with a constant gain $\xi = 0.8$. The secondary peak in $|G(\omega)|^2$ exceeds 1. Consequently the vehicles of the platoon have increasingly strong oscillations in acceleration as $n$ increases. For example, the acceleration of the 50$^{th}$ car is displayed in Fig. 9. In stark contrast, no oscillations are observed when the frequency-dependent feedback is used.

Finally, Fig. 10 depicts the acceleration of vehicles with the frequency-dependent feedback. The improvement in smoothness (lower maximum acceleration) is demonstrated by a comparison to Fig. 4.

## 6. Conclusions

This paper presents a methodology to calculate the dynamic response of a platoon of adaptive cruise control vehicles given the mechanical response function of an individual car $g(t)$. The control algorithm considered is based on a constant headway time, relative velocity feedback, and acceleration feedback. Simulations were performed using a model $g(t)$ that approximates the measured torque response of an engine (Stefanopoulou and Kolmanovsky, 1999). The behavior of vehicles following a leading vehicle that accelerates at 1 m/s$^2$ from 5 m/s to 10 m/s has been calculated. For sufficiently large acceleration-feedback gain $\xi$, the platoon is string stable and the peak acceleration decreases with increasing car number $n$. An asymptotically accurate approximation for the dependence on $n$ is given by $erf\left(\frac{ct.}{\sqrt{n}}\right)$. In addition to the main peak in acceleration, oscillations of period 2-3 s can occur for small gain $\xi$ that marginally produces string stability. However, a benefit of small gain is a quicker decrease of peak acceleration with increasing car number and thus a smoother ride. It was demonstrated that a frequency-dependent gain produces better performance than a constant gain, but at the expense of additional computation and memory.

**References**

Bareket, Z., Fancher, P. S., Peng, H., Lee, K., Assaf, C. A., 2003. Methodology for assessing cruise control behavior. IEEE Transactions on Intelligent Transportation Systems 4 (3), 123-131.

Bose, A., Ioannou, P., 2003a. Mixed manual/semi-automated traffic: a macroscopic analysis. Transportation Research Part C 11, 439-462.

Bose, A., Ioannou, P., 2003b. Analysis of Traffic Flow With Mixed Manual and Semiautomated Vehicles. IEEE Transactions on Intelligent Transportation Systems 4 (4), 173-187.

Davis, L. C., 2004. Effect of adaptive cruise control systems on traffic flow. Physical Review E 69, 066110.

Davis, L. C., 2007. Effect of adaptive cruise control systems on mixed traffic flow near an on-ramp. Physica A 379, 274-290.

Davis, L. C., 2012. Stability of adaptive cruise control systems taking account of vehicle response time and delay. Physics Letters A 376, 2658-2662.

Davis, L. C., 2013. Optimality and oscillations near the edge of stability in the dynamics of autonomous vehicle platoons. Submitted to Physica A.

Ioannou, P., Chien, C. C., 1993. Autonomous intelligent cruise control. IEEE Transactions on Vehicular Technology 42 (4), 657-671.

Ioannou, P. A., Stefanovic, M., 2005. Evaluation of ACC vehicles in mixed traffic: lane change effects and sensitivity analysis. IEEE Transactions on Intelligent Transportation Systems 6 (1), 79-89.

Jiang, R., Wu, Q.-S., 2006. The adaptive cruise control vehicles in the cellular automata model. Physics Letters A 359, 99-102.

Jiang, R., Hu, B., Jia, M.-B., Wang, R., Wu, Q.-S., 2007. Phase transition in a mixture of adaptive cruise control vehicles and manual vehicles. The European Physical Journal B - Condensed Matter and Complex Systems 58, 197-206.

Kerner, B. S., in: H.S. Mahmassani (Ed.), Transportation and Traffic Theory Flow, Dynamics and Human Interactions, Proceedings of the 15th International Symposium on Transportation and Traffic Theory, Elsevier, Boston, 2005, p. 198.

Kesting, A., Treiber, M., Schönhof, M., Helbing, D., 2008. Adaptive cruise control design for active congestion avoidance. Transportation Research Part C 16, 668–683.

Li, P. Y., Shrivastava, A. 2002. Traffic flow stability induced by constant time headway policy for adaptive cruise control vehicles. Transportation Research Part C 10, 275–301.

Liang, C.-Y., Peng, H., 1999. Optimal Adaptive Cruise Control with Guaranteed String Stability. Vehicle System Dynamics 32, 313–330.

Liang, C.-Y., Peng, H., 2000. String stability analysis of adaptive cruise controlled vehicles. JSME International Journal Series C: Mechanical Systems, Machine Elements and Manufacturing 43, 671-677.

Orosz, G., Moehlis, J., Bullo, F., 2010. Robotic reactions: Delay-induced patterns in autonomous vehicle systems. Physical Review E 81, 025204(R).

Orosz, G., Moehlis, J., Bullo, F., 2011. Delayed car-following dynamics for human and robotic drivers in: Proceedings of the ASME 2011International Design Engineering Technical Conferences & Computers and Information in Engineering Conference IDETC/CIE 2011, Washington, DC, USA, August 28–31, 2011.

Sheikholeslam, S., Desoer, C. A., 1990. Longitudinal control of a platoon of vehicles. In Proceedings of American Control Conference, May 1990, pp. 291-296.

Stefanopoulou, A., Kolmannovsky, I., 1999. Analysis and control of transient torque response in engines with internal exhaust gas recirculation. IEEE Transactions on Control Systems Technology 7, (5), 555-566.

Swaroop, D., Hedrick, J. K., Chien, C. C., Ioannou, P. A., 1994. A Comparision of Spacing and Headway Control Laws for Automatically Controlled Vehicles. Vehicle System Dynamics 23, 597-625.

Swaroop, D., Hedrick, J.K., 1996. String stability of interconnected systems. IEEE Transactions on Automatic Control 41, 349-357.

Swaroop, D., Rajagopal, K.R., 1999. Intelligent cruise control systems and traffic flow stability. Transportation Research Part C 7, 329-352.

Treiber, M., Helbing, D., 2001. Microsimulations of freeway traffic including control measures. Automatisierungstechnik 49, 478-484.

VanderWerf, J., Shladover, S. E., Miller, M. A., Kourjanskaia, N., 2002. Effects of adaptive cruise control systems on highway traffic flow capacity. Transportation Research Record, no. 1800, 78-84.

Yi, J., Horowitz, R., 2006. Macroscopic traffic flow propagation stability for adaptive cruise controlled vehicles. Transportation Research Part C 14, 81-95.

Yuan, Y.-M., Jiang, R., Hu, M.-B., Wu, Q.-S., Wang, R., 2009. Traffic flow characteristics in a mixed traffic system consisting of ACC vehicles and manual vehicles: A hybrid modelling approach. Physica A 388, 2483-2491.

Zhou, J., Peng, H., 2005. Range Policy of Adaptive Cruise Control Vehicles for Improved Flow Stability and String Stability. IEEE Transactions on Intelligent Transportation Systems 6 (2), 229-237.

**Table 1.** Values of the parameters in the control algorithm.

| | |
|---|---|
| $\alpha$ | 2 s$^{-1}$ |
| $h$ | 1 s |
| $D$ | 5 m |
| $k$ | 1 s$^{-1}$ |

**Figure captions**

Fig. 1. A response function adapted from the "fast" torque curve at 750 rpm reported by A. G. Stefanopoulou and I. Kolmanovsky (Fig. 9 of Stefanopoulou and Kolmannovsky, 1999).

Fig. 2. The absolute square of the transfer function vs. angular frequency for two acceleration-feedback gains $\xi = 0.82$ and $0.9$.

Fig. 3. The acceleration of vehicles (1-50) following a leading vehicle that accelerates at $1$ m/s$^2$ for 5 s. The velocity goes from 5 m/s to 10 m/s. Oscillations occurring after main peak are due to the secondary peak in $|G(\omega)|^2$ at $\omega \approx 2.6$ rad/s for $\xi = 0.82$.

Fig. 4. The acceleration of vehicles (1-50) following a leading vehicle that accelerates as in Fig. 3. The acceleration-feedback gain is $\xi = 0.9$. Oscillations observed in Fig. 3 are nearly absent because the secondary peak in $|G(\omega)|^2$ is weaker.

Fig. 5. The peak (maximum) acceleration vs. car number $n$ (blue diamonds) for $\xi = 0.9$. The red curve is the approximation given by Eq. (20b), $a^{peak} \sim \text{erf}\left(\frac{ct.}{\sqrt{n}}\right), n \to \infty$.

Fig. 6. The transform in the time domain $\tilde{\xi}_p(t)$ of the frequency-dependent portion of the gain in a frequency-dependent acceleration-feedback algorithm with $\xi_p = 0.3$ and $\omega_p = 5$ rad/s. See Eqs. (25) and (26).

Fig. 7. The absolute square of the transfer function, $|G(\omega)|^2$ (blue curve), with a frequency-dependent acceleration gain, $\xi(\omega)$ (red curve), with $\xi_0 = 0.8, \xi_p = 0.3$ and $\omega_p = 5$ rad/s. See Eq. (22).

Fig. 8. The absolute square of the transfer function, $|G(\omega)|^2$ (blue curve), with a constant gain $\xi = 0.8$ (red line). The secondary peak in $|G(\omega)|^2$ exceeds 1.

Fig. 9. The acceleration of the 50$^{th}$ car with the same leading–vehicle acceleration as in Fig. 3. The blue curve is for the frequency-dependent acceleration-feedback gain with $\xi_0 = 0.8, \xi_p = 0.3$ and $\omega_p = 5$ rad/s. The red curve with strong oscillations corresponds to the transfer function in Fig. 8.

Fig. 10. The acceleration of vehicles (1-50) with the frequency-dependent acceleration-feedback gain of Fig. 7 and the same leading–vehicle acceleration as in Fig. 3.

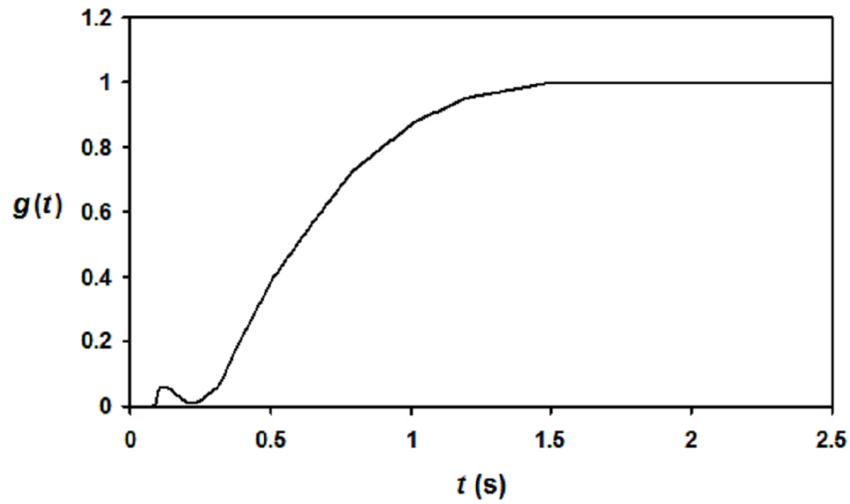

Fig. 1. A response function adapted from the "fast" torque curve at 750 rpm reported by A. G. Stefanopoulou and I. Kolmanovsky (Fig. 9 of Stefanopoulou and Kolmannovsky, 1999).

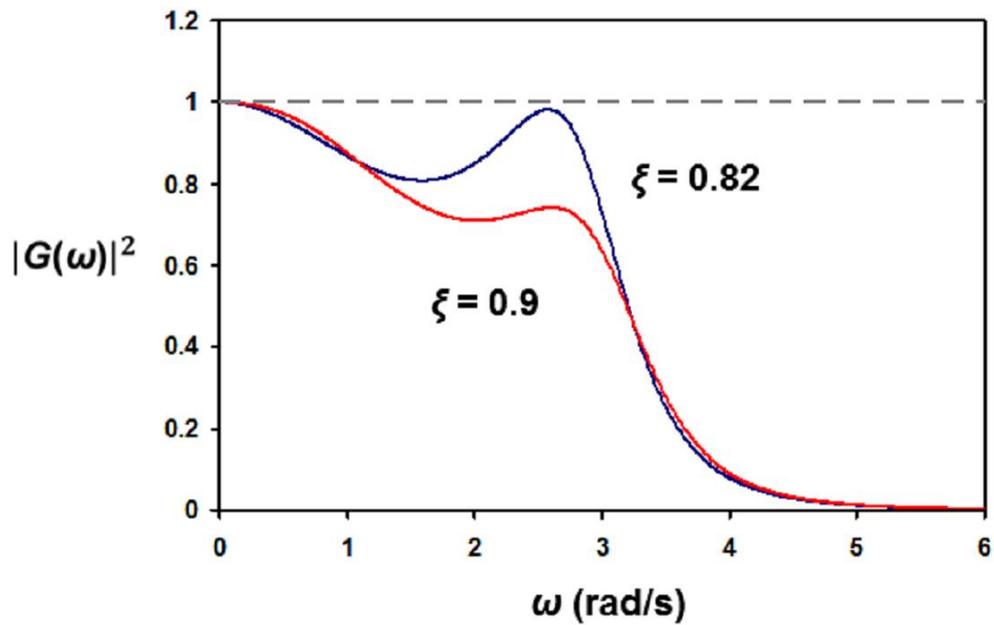

Fig. 2. The absolute square of the transfer function vs. angular frequency for two acceleration-feedback gains $\xi = 0.82$ and $0.9$.

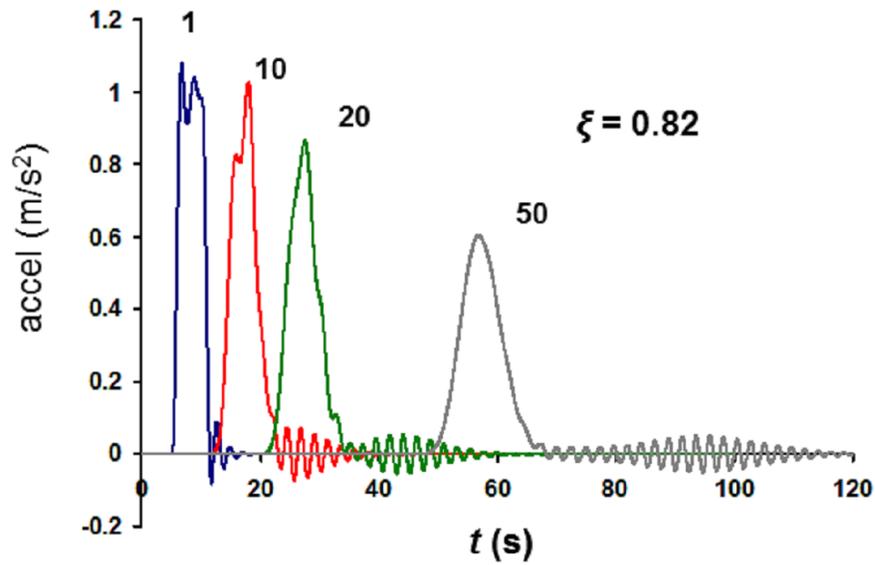

Fig. 3. The acceleration of vehicles (1-50) following a leading vehicle that accelerates at 1 m/s² for 5 s. The velocity goes from 5 m/s to 10 m/s. Oscillations occurring after main peak are due to the secondary peak in $|G(\omega)|^2$ at $\omega \approx 2.6$ rad/s for $\xi = 0.82$.

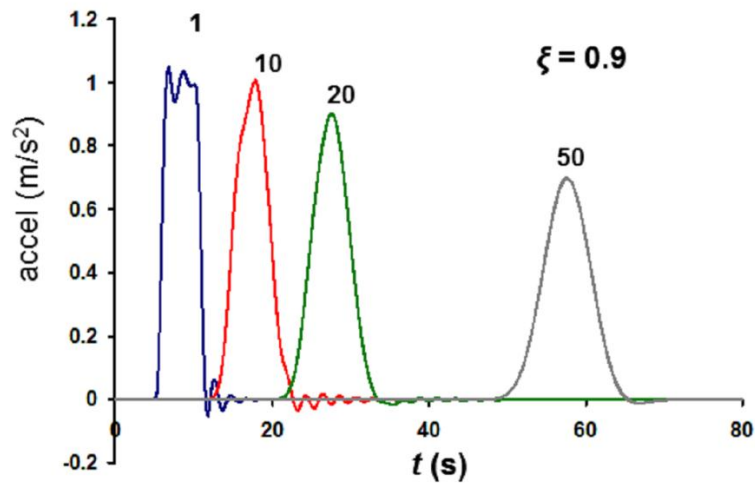

Fig. 4. The acceleration of vehicles (1-50) following a leading vehicle that accelerates as in Fig. 3. The acceleration-feedback gain is $\xi = 0.9$. Oscillations observed in Fig. 3 are nearly absent because the secondary peak in $|G(\omega)|^2$ is weaker.

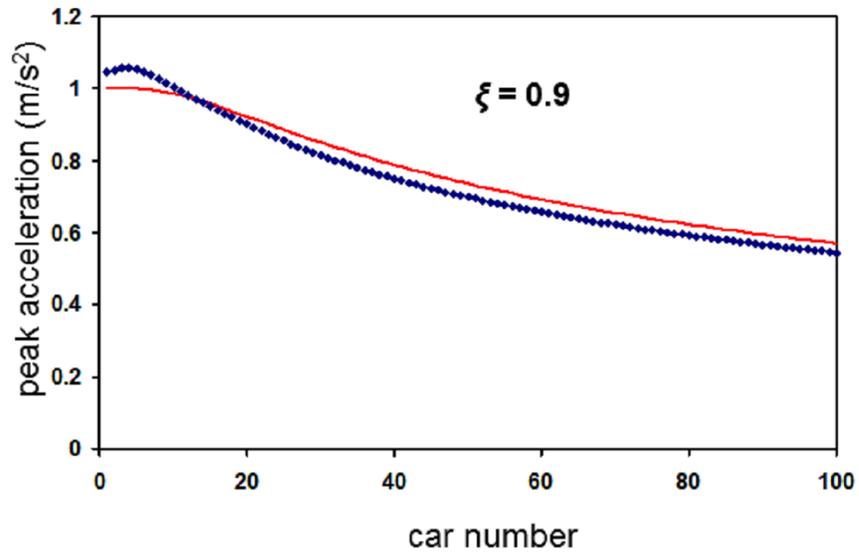

Fig. 5. The peak (maximum) acceleration vs. car number $n$ (blue diamonds) for $\xi = 0.9$. The red curve is the approximation given by Eq. (20b), $a^{peak} \sim \text{erf}\left(\frac{ct.}{\sqrt{n}}\right), n \to \infty$.

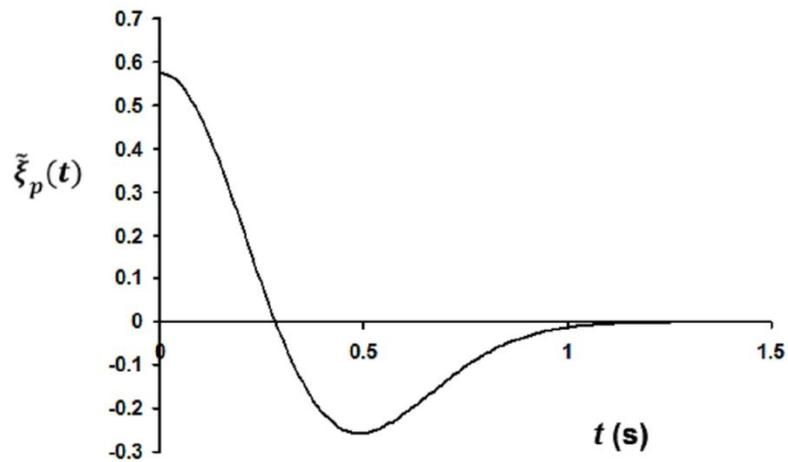

Fig. 6. The transform in the time domain $\tilde{\xi}_p(t)$ of the frequency-dependent portion of the gain in a frequency-dependent acceleration-feedback algorithm with $\xi_p = 0.3$ and $\omega_p = 5$ rad/s. See Eqs. (25) and (26).

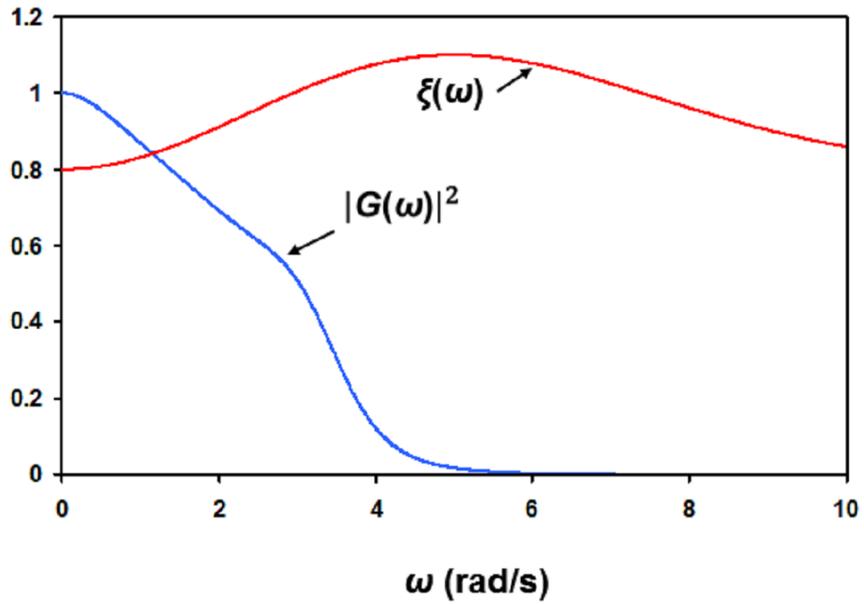

Fig. 7. The absolute square of the transfer function, $|G(\omega)|^2$ (blue curve), with a frequency-dependent acceleration gain, $\xi(\omega)$ (red curve), with $\xi_0 = 0.8, \xi_p = 0.3$ and $\omega_p = 5$ rad/s. See Eq. (22).

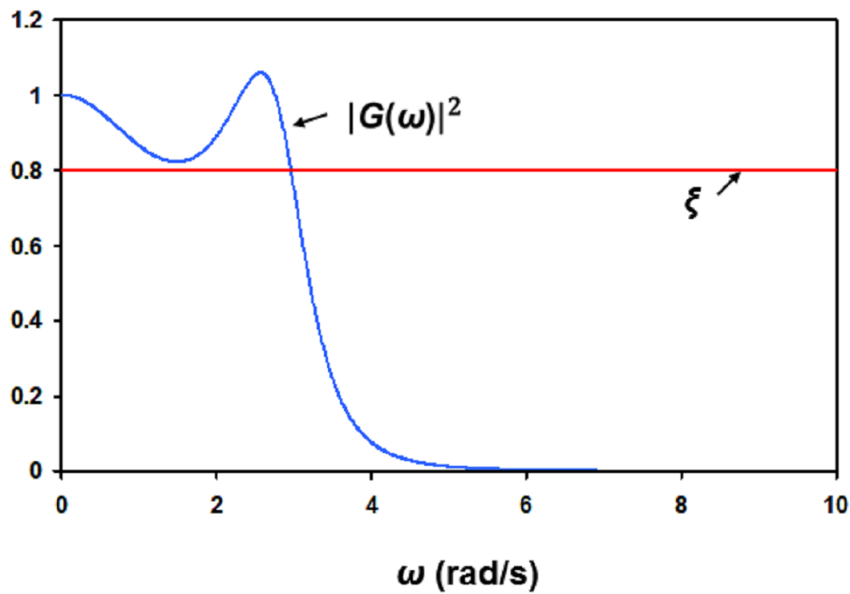

Fig. 8. The absolute square of the transfer function, $|G(\omega)|^2$ (blue curve), with a constant gain $\xi = 0.8$ (red line). The secondary peak in $|G(\omega)|^2$ exceeds 1.

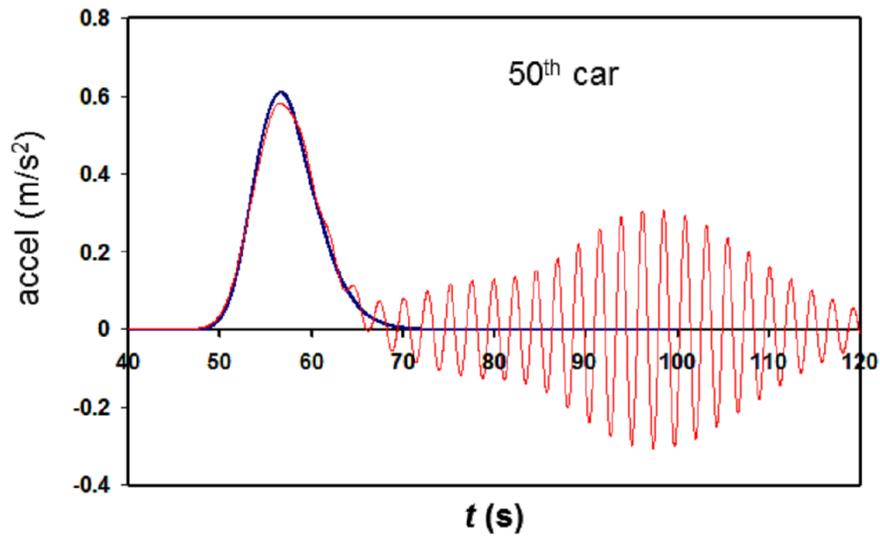

Fig. 9. The acceleration of the 50[th] car with the same leading–vehicle acceleration as in Fig. 3. The blue curve is for the frequency-dependent acceleration-feedback gain with $\xi_0 = 0.8, \xi_p = 0.3$ and $\omega_p = 5$ rad/s. The red curve with strong oscillations corresponds to the transfer function in Fig. 8.

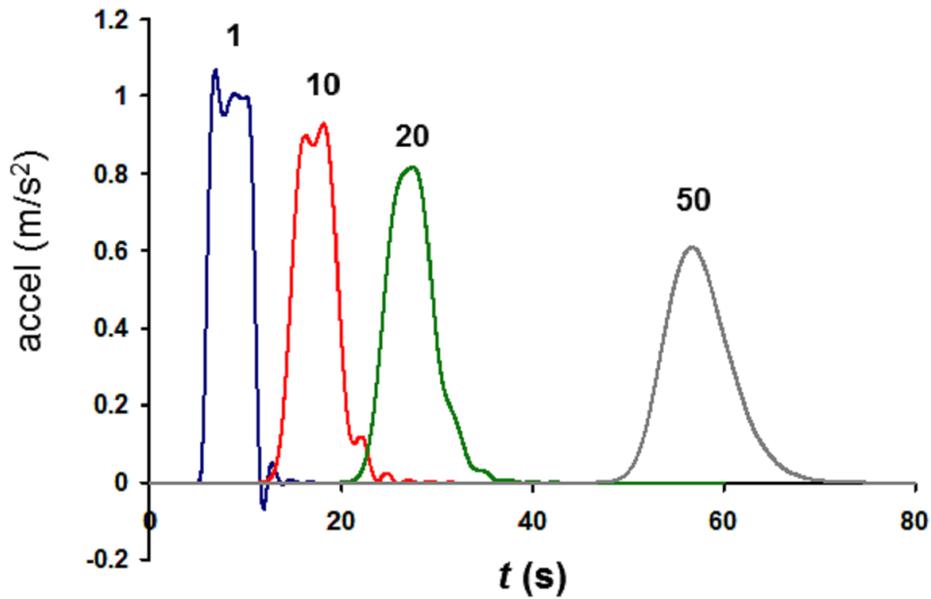

Fig. 10. The acceleration of vehicles (1-50) with the frequency-dependent acceleration-feedback gain of Fig. 7 and the same leading–vehicle acceleration as in Fig. 3.